\documentstyle[12pt,a4,epsf]{article}
\begin{document} 
\bibliographystyle{unsrt} 
\baselineskip= 18pt 
\pagenumbering{arabic} 
\pagestyle{plain} 
\def \bit{\begin{itemize}}
\def \eit{\end{itemize}}
\def \half{{1\over 2}}
\def \OO{\Omega}
\def \aa{\alpha}
\def \ba{{\bf a}}
\def \bk{{\bf k}}
\def \bkp{{\bf k'}}
\def \bqp{{\bf q'}}
\def \bq {{\bf q}}
\def \bpp{{\bf p'}}
\def \bp {{\bf p}}
\def \EE{\Bbb E}
\def \EEx{\Bbb E^x}
\def \EEo{\Bbb E^0}
\def \LL{\Lambda}
\def \PP{\Bbb P^o}
\def \rr{\rho}
\def \SS{\Sigma}
\def \ss{\sigma}
\def \ll{\lambda}
\def \dd{\delta} \def \ee{\epsilon} \def \ww{\omega}
\def \bww{\mbox{\boldmath $\omega$}}
\def \DD{\Delta}
\def \DDt{\tilde {\Delta}}
\def \kr{\kappa\lb \LL\rb}
\def \PPx{\Bbb P^{x}}
\def \gg{\gamma}
\def \GG{\Gamma}
\def \kk{\kappa}
\def \tt{\theta}
\def \lb{\left(}
\def \la{\langle}
\def \ra{\rangle}
\def \rb{\right)}
\def \prt{\tilde p}
\def \pt{\tilde {\phi}}
\def \bb{\beta}
\def \hal{{1\over 2}\nabla ^2}
\def \bg{{\bf g}}
\def \bx{{\bf x}}
\def \bl{{\bf l}}
\def \bu{{\bf u}}
\def \bv{{\bf v}}
\def \by{{\bf y}}
\def \bw{{\bf w}}
\def \bU{{\bf U}}
\def \bR{{\bf R}}
\def \hag{{1\over 2}\nabla}
\def \beq{\begin{equation}}
\def \eeq{\end{equation}}
\def \bea{\begin{eqnarray}}
\def \eea{\end{eqnarray}}
\def \cosech{\hbox{cosech}}
\def \fr{\frac}
\def \d{\partial}
\def \Gc{{\cal G}}
\def \dmq{\fr{d^D\bq}{(2\pi)^D}}
\def \dmqp{\fr{d^D\bq'}{(2\pi)^D}}
\def \dmk{\fr{d^D\bk}{(2\pi)^D}}
\def \dmkp{\fr{d^D\bk'}{(2\pi)^D}}
\def \dmp{\fr{d^D\bp}{(2\pi)^D}}
\def \dmpp{\fr{d^D\bp'}{(2\pi)^D}}
\def \Tr{{\mbox{Tr}}}

\title{Continuum Derrida Approach to Drift and Diffusivity in Random Media}
\author{D S Dean$^{\dag}$, I T Drummond and R R Horgan  \\
        Department of Applied Mathematics and Theoretical Physics \\
        University of Cambridge \\
        Silver St \\
        Cambridge, England CB3 9EW\\~~\\~~\\
        ${~^{\dag}}$Dipartimento di Fisica and INFN\\
        Univesit\`a di Roma La Sapienza \\
        P. A. Moro 2, 00185 Roma, Italy}
\maketitle

\begin{abstract}

By means of rather general arguments, based on
an approach due to Derrida that makes use of samples of finite size,
we analyse the effective diffusivity and drift tensors
in certain types of random medium in which the
motion of the particles is controlled by molecular diffusion and 
a local flow field with known statistical properties.  

The power of the Derrida method is that it uses the
equilibrium probability distribution,
that exists for each {\em finite} sample, to compute asymptotic behaviour
at large times in the {\em infinite} medium.
In certain cases, where this equilibrium situation is associated with
a vanishing microcurrent, our results demonstrate the 
equality of the renormalization processes for the effective
drift and diffusivity tensors. This establishes, for those cases,
a Ward identity previously verified only to two-loop order 
in perturbation theory in certain models.

The technique can be applied also to media in which the diffusivity
exhibits spatial fluctuations. We derive a simple relationship between
the effective diffusivity in this case and that for an associated
gradient drift problem that provides an interesting constraint on
previously conjectured results.
\end{abstract}
\vfill
DAMTP-96-80
\pagebreak
 
\section{Introduction}

Problems of flow and diffusion in random environments are
of great interest for many reasons both mathematical and
physical \cite{jpmega,RUSS1,RUSS2}. Possible applications 
range from turbulent diffusion \cite{kraichnan}-\cite{itdetal3}
to ionic and molecular diffusion in the presence of trapped ions,
dislocations and other impurities \cite{GLAD}.

A typical continuum model, that represents some aspects
of these physical systems, comprises a particle in a $D$ dimensional space
subject to
molecular diffusion, $\kk_{ij}$, together with a local drift 
velocity, $\bu(\bx)$~. The motion of a particle at position $\bx$
is given by
\beq
\dot\bx=\bu(\bx)+\bw(t)~~,
\label{MOTION}
\eeq
where $\bw(t)$ is a white noise process that satisfies
\beq
\la w_i(t)\ra=0~~~~~~~~
\mbox{and}~~~~~~~~\la w_i(t)w_j(t')\ra=2\kk_{ij}\delta(t-t')~~.
\eeq
The flow field $\bu(\bx)$ is time independent and 
assumed to have certain statistical 
properties, the details of which depend on the particular model.
However we do assume that the system is homogeneous so that, for example, 
\beq
\la u_i(\bx)u_j(\bx')\ra=\Delta_{ij}(\bx-\bx')~~,
\eeq
and $\la\bu(\bx)\ra$ is independent of position. 
Frequently the statistics of the velocity field are assumed (for simplicity)
to be Gaussian in character so that only the mean velocity and its two-point 
correlator are required to determine all correlators. We use angle brackets
to indicate an average over the white noise ensemble and/or over the
ensemble of samples of the medium as appropriate. The intended version of the averaging
procedure will be clear from the context.

The computational problem we address then, is to use the above 
statistical information on the 
model to deduce the effective drift and diffusivity that control the
motion of the particles at large distances and times - the bulk 
properties of the medium. There has been a 
great deal of work on this problem \cite{kraichnan}-\cite{itdetal3},
\cite{mythesis}-\cite{USDIF4}. 
In this paper we develop a continuum
version of an approach by
Derrida \cite{jpmega,DERRIDA} that he applied to lattice models 
in order to elucidate relationships
that may exist between the effective long range parameters of the theory,
one example of which is the ``Einstein relation'' between drift and 
diffusivity \cite{RUSS1,RUSS2}.

\section{Derrida Argument for the Continuum}

The idea behind Derrida's argument \cite{DERRIDA}, which is drawn from statistical physics,
is that the bulk properties of an {\em infinite} medium
can be captured in a sample of {\em finite} size provided this size is
much larger than the typical correlation lengths of 
the statistical fluctuations of the velocity field $\bu(\bx)$~.
The size of the sample provides an infra-red cut-off, to use the language
of field theory. The assumption being made is that the effective
parameters are insensitive to the infra-red cut-off in a limit in which it 
becomes very large. Of course the presence of important long
range correlations would lead to a dependence of the effective parameters 
on the sample size (and geometry). These are circumstances in which we would
expect to see anomalous diffusion \cite{jpmega}. 

Another important issue in relation to the
removal of the infra-red cut-off is the extent to which this limit commutes
with the large time limit invoked in the definition of bulk quantities \cite{jpmega}~.
We do not address this point mathematically in this paper but rely,
for the moment, on physical intuition. 
The commutation of these limits is a very 
important issue both from a mathematical and a physical point of view
and should be examined more thoroughly.

We will assume, therefore, that the medium is represented by a velocity field
$u_i(\bx)$ that is defined over a basis region $\Omega$
and that space is tessellated by this region under displacements $\bl$
that lie in a Bravais lattice $B$~. For the purposes of visualization it is
sufficient to suppose that $\Omega$ is cube of side $L$~. The velocity field $u_i(\bx)$  
then satisfies
$$
u_i(\bx+\bl)=u_i(\bx)~~,
$$
and we assume that $\Omega$ which contains the origin, is very large
relative to the basic correlation length in $u_i(\bx)$.

A blob of particles is released from the origin at $t=0$~. They move independently
according to eq(\ref{MOTION}) and are free to cross the boundaries of
$\Omega$ and it's replicates.
The resulting probability distribution, $P(\bx,t)$,
satisfies
\beq
\fr{\d}{\d t}P(\bx,t)=\d_i(\kk_{ij}\d_j-u_i(\bx))P(\bx,t)~~.
\eeq

All the moments of the distribution can be obtained from the 
moment generating function of the distribution 
\beq
Z(\bk)=\int d^D\bx e^{-i\bk.\bx}P(\bx,t)
\eeq
The technical step suggested by Derrida is to re-write this integral
over all space as a sum over replicates of the basis region $\Omega$,
thus
\beq
Z(\bk)=\int_{\Omega}d^D\bx \sum_{\bl} e^{-i\bk.(\bx+\bl)}P(\bx+\bl,t)~~.
\eeq
That is
\beq
Z(\bk)=\int_{\Omega}d^D\bx W(\bx,t)=\int_{\Omega}d^D\bx e^{-i\bk.\bx}R(\bx,t)~~,
\eeq
where
\beq
R(\bk,\bx,t)=\sum_{\bl} e^{-i\bk.\bl}P(\bx+\bl,t)~~,
\eeq
and
\beq
W(\bk,\bx,t)=e^{-i\bk.\bx}R(\bk,\bx,t)~~,
\eeq
The probability distribution can be reconstructed as
\beq
P(\bx,t)=\int\fr{d^D\bk}{(2\pi)^D}e^{i\bk.\bx}Z(\bk)~~.
\eeq
Note that there is no restriction on the integration range of $\bk$~.

Making use of the periodicity properties of the velocity field
we can show that 
\beq
\fr{\d}{\d t}R(\bk,\bx,t)=\d_i(\kk_{ij}\d_j-u_i(\bx))R(\bk,\bx,t)~,
\eeq
where $R(\bx,t)$ satisfies the boundary condition
\beq
R(\bk,\bx+\ba,t)=e^{i\bk.\ba}R(\bk,\bx,t)~~,
\eeq
with $\ba$ an element of the Bravais lattice.

As $t\rightarrow\infty$ we expect that
time dependence will disappear.
In that limit
\beq
\d_i(\kk_{ij}\d_j-u_i(\bx))R(\bk,\bx,t)=0~~.
\eeq
This implies that for large $t$
\beq
R(\bk,\bx,t)\propto Q_0(\bx)
\eeq
where $Q_0(\bx)$ is the static probability distribution in $\Omega$~.
The advantage of the Derrida method is precisely that one may make use of this static
equilibrium distribution to evaluate asymptotic behaviour.

However $Q_0(\bx)$ is periodic under displacements in the Bravais lattice and, except
for the case $\bk=0$, $R(\bk,\bx,t)$ is {\it not} periodic. Hence
we must have
\beq
R(0,\bx,t)\rightarrow Q_0(\bx)~~,
\eeq
and
\beq
R(\bk,\bx,t)\rightarrow 0~~.
\eeq
when $\bk\ne 0$~. All of this is entirely consistent with 
continuity in $\bk$ at finite time.

The equation satisfied by $W(\bk,\bx,t)$, which {\it is} periodic on the Bravais lattice, is
\beq
\fr{\d}{\d t}W(\bk,\bx,t)=(\d_i+ik_i)(\kk_{ij}(\d_j+ik_j)-u_i(\bx))W(\bk,\bx,t)~~.
\label{DEREQ}
\eeq

When $\bk=0$ we know that the operator on the right has an eigenfunction 
with a zero eigenvalue, namely $Q_0(\bx)$, the stationary probability
distribution. When $\bk\ne 0$ this eigenfunction is perturbed and
the eigenvalue is moved away from zero. We denote the eigenfunction by $Q(\bx)$
and the eigenvalue by $-\mu(\bk)$ so that $\mu(0)=0$ and
\beq
(\d_i+ik_i)(\kk_{ij}(\d_j+ik_j)-u_i(\bx))Q(\bx)=-\mu(\bk)Q(\bx)~~.
\label{DEREIG}
\eeq
An examination of the (trivial) case with {\em constant} velocity field
suggests that near $\bk=0$ there is a gap in the spectrum of this operator
of magnitude $\sim (2\pi/L)^{2}$ provided $|\bk|$ much less than $2\pi/L$~. 
We assume this to be true
and conclude that up to exponentially damped corrections that are $O(e^{-(2\pi/L)^2t})$,
the solution of eq(\ref{DEREQ}) in which we are interested is
\beq
W(\bk,\bx,t)=Q(\bx)e^{-\mu(\bk)t}~~.
\eeq
We now impose the normalization condition
\beq
\int_{\Omega}d^D\bx Q(\bx)=1~~,
\label{DERNORM}
\eeq
to obtain the resulting formula for the moment generating function is
\beq
Z(\bk)=e^{-\mu(\bk)t}~~,
\eeq
where we have taken into account eq(\ref{DERNORM})~.
The corresponding formula for the probability distribution in
the unbounded (but tessellated) space is therefore
\beq
P(\bx,t)=\int\dmk e^{i\bk.\bx-\mu(\bk)t}~~.
\label{PROBDIST}
\eeq
For large $t$ the asymptotic form of $P(\bx,t)$ is determined 
by the terms in $\mu(\bk)$ up to $O(\bk^2)$~. The remaining 
terms of higher order determine the approach to this asymptotic form in 
inverse powers of $\sqrt{t}$ in the general case. 

It should be noted that on the basis of the above argument the asymptotic
behaviour is guaranteed to set in only when $t$ is so large that only
values of $|\bk|$ much less than $2\pi/L$ yield important contributions to the integral
in eq(\ref{PROBDIST})~. This means that we are computing asymptotic behaviour
that becomes apparent after the probability distribution of the particle has spread over
more than one replicate of $\Omega$~. It is the physical assumption of the Derrida 
method that, when the samples are sufficiently large and there is no anomalous
diffusive behaviour, this asymptotic behaviour sets in effectively much
earlier and therefore there is no change to the estimated bulk diffusivity
and drift when the infra-red cutoff is removed. Proving this mathematically
remains to be done. Any such proof should also tell us about the physically important
issue of how effects on different scales in the problem interact with one another.

\section{Asymptotic Behaviour}

We can calculate these terms in $\mu(\bk)$
by perturbation theory. Set
\beq
Q(\bx)=Q_0(\bx)+Q_1(\bx)+Q_2(\bx)+\cdots~~,
\eeq
and 
\beq
\mu(\bk)=\mu_1(\bk)+\mu_2(\bk)+\cdots~~,
\eeq
where the suffix on a term indicates its order in powers of $\bk$~.
Note that there is no term $\mu_0$ and that the normalization of
$Q_0(\bx)$ means that
\beq
\int d^D\bx Q_1(\bx)=\int d^D\bx Q_2(\bx)=0~~.
\eeq
This means that $Q_1$, $Q_2$ and all higher corrections
lie in what we will refer to as the zero weight subspace. We can obtain the equations
for the higher corrections by expanding both sides of eq(\ref{DEREIG})
in powers of $\bk$~. We find
\beq
\d_l(\kk_{lm}\d_m-u_l(\bx))Q_1(\bx)+ik_l(2\kk_{lm}\d_m-u_l(\bx))Q_0(\bx)
                                  =-\mu_1(\bk)Q_0(\bx)~~,
\label{PERT1}
\eeq
and
\beq
\d_l(\kk_{lm}\d_m-u_l(\bx))Q_2(\bx)+ik_l(2\kk_{lm}\d_m-u_l
(\bx))Q_1(\bx)-k_lk_m\kk_{lm}Q_0(x)
~~~~~~~~~~~~~~~~~~~~~~~~~~~~~~~~
\label{PERT2}
\eeq
$$
~~~~~~~~~~~~~~~~~~~~~~~~~~~~~~~~~~
                                  =-\mu_2(\bk)Q_0(\bx)-\mu_1(\bk)Q_1(\bx)~~.
$$
For completeness we note that the general recursion formula ($n\ne 2$)
is
\beq
\d_l(\kk_{lm}\d_m-u_l(\bx))Q_n(\bx)+ik_l(2\kk_{lm}\d_m-u_l(\bx))Q_{n-1}(\bx)
~~~~~~~~~~~~~~~~~~~~~~~~~~~~~~~~~~~~~~~~~~~~~~~
\label{PERT3}
\eeq
$$
~~~~~~~~~
                 =-\mu_n(\bk)Q_0(\bx)-\cdots 
         -(\mu_2(\bk)-k_lk_m\kk_{lm})Q_{n-2}(\bx)-\mu_1(\bk)Q_{n-1}(\bx)~~.
$$

After integration over the base cell $\Omega$, eq(\ref{PERT1}) yields
\beq
\mu_1(\bk)=ik_l\int_{\Omega} d^D\bx u_l(\bx)Q_0(\bx)=ik_l\bar{u}_l
\label{MEANVEL}
\eeq
where $\bar{u}_l$ is the average value of the velocity in the stationary
probability distribution $Q_0(\bx)$~. Note that the vanishing, 
under integration, of the gradient 
terms is assured by the periodicity and continuity of the random
field.

To compute the next correction we need the Green's function $G(\bx,\bx')$ 
that is the inverse of
the operator 
$$
H=\d_l(\kk_{lm}\d_m-u_l(\bx))
$$ 
in the zero weight subspace. The existence of this Green's function depends
on our physically plausible assumption that the kernel of $H$ is 
one-dimensional and is therefore spanned by $Q_0(\bx)$~.

We have then
\beq
\d_l(\kk_{lm}\d_m-u_l(\bx))G(\bx,\bx')=-\delta(\bx,\bx')
                     \equiv -\delta(\bx-\bx')+\fr{1}{V}~~,
\eeq
where $V$ is the volume of $\Omega$~.
It follows then that
\beq
Q_1(\bx)=ik_l\int_{\Omega}d^D\bx' 
                G(\bx,\bx')(2\kk_{lm}\d'_m-(u_l(\bx')-\bar{u_l}))Q_0(\bx')
\eeq
The evaluation of $\mu_2$ is achieved by integrating eq(\ref{PERT2})
with respect to $\bx$ over $\Omega$~.
We have
\beq
\mu_2=k_l\kk_{lm}k_m+ik_l\int_{\Omega}d^D\bx u_l(\bx)Q_1(\bx)
\eeq
That is
\beq
\mu_2=k_l\kk_{lm}k_m-k_lk_m\int_{\Omega}d^D\bx d^D\bx' 
                    u_l(\bx)G(\bx,\bx')(2\kk_{mn}\d'_n-(u_m(\bx')-\bar{u}_m))Q_0(\bx')
\eeq

These results tell us that the mean velocity of the cloud is $\bar{u}_i$
and the effective diffusivity tensor is
\beq
\kk^{\rm{e}}_{lm}=\kk_{lm}-\fr{1}{2}\left\{\int_{\Omega}d^D\bx d^D\bx'
                       \tilde{u}_l(\bx)G(\bx,\bx')(2\kk_{mn}\d'_n-\tilde{u}_m(\bx'))Q_0(\bx')
                         +(l\leftrightarrow m)\right\}
\label{EFFDIFF0}
\eeq
where we have set
\beq
\tilde{u}_l(\bx)=u_l(\bx)-\bar{u}_l(\bx)~~,
\eeq
and used the result
\beq
\int_{\Omega}d^D\bx G(\bx,\bx')=0~~.
\eeq

\section{Gradient Flow}

A case of particular interest is that of gradient flow where
the local drift depends linearly on the gradient of a scalar field.
Thus
\beq
u_l(\bx)=\ll_{lm}\d_m\phi(\bx)~~.
\eeq

In this case the equilibrium microcurrent is
\beq
J_l(\bx)=-(\kk_{lm}\d_m-\ll_{lm}\d_m\phi(\bx))Q_0(\bx)~~.
\eeq
Within this class of models an important case arises
for which the drift and diffusivity tensors are proportional
to one another,
\beq
\ll_{lm}=\tau\kk_{lm}~~.
\eeq
In this case the microcurrent vanishes in equilibrium
and the static probability distribution satisfies
\beq
\d_l\kk_{lm}(\d_l-\tau\d_l\phi(\bx))Q_0(\bx)=0~~.
\label{STATIC1}
\eeq
That is 
\beq
Q_0(\bx)=Ne^{\tau\phi(\bx)}~~,
\eeq
where N is a normalization constant. 
It is also the case that
\beq
\bar{u}_l(\bx)=\int_{\Omega} d^D\bx\tau \kk_{lm}\d_m\phi(\bx) Q_0(\bx)=0~~.
\eeq
It follows that $\tilde{\bf u}(\bx)=\bu(\bx)$~.

The formula for the effective diffusivity, eq(\ref{EFFDIFF0}), becomes
\beq
\kk^{\rm{e}}_{lm}=\kk_{lm}-\int_{\Omega}d^D\bx d^D\bx'
                        u_l(\bx)G(\bx,\bx')u_m(\bx')Q_0(\bx')
\label{EFFDIFF1}
\eeq
which is symmetrical in $l$ and $m$~.                
Since $G(\bx,\bx')Q_0(\bx')$ is a positive operator we see that the tendency is
for the random behaviour of the medium to reduce the effective diffusivity in 
a given direction relative to the microscopic value. This is consistent with
the idea that the random fluctuations of the field $\phi(\bx)$ represent
a trapping mechanism tending to hold the particle near the bottoms of 
potential wells.

\section{External Gradient}

To investigate the computation of the effective drift we add a
constant external gradient to the fluctuating gradient. We have
\beq
u_l(\bx)=\tau\kk_{lm}(\d_m\phi(\bx)+g_m)~~.
\eeq
The static probability distribution now satisfies
\beq
\d_l\kk_{lm}(\d_m-\tau(\d_m\phi(\bx)+g_m))Q^g_0(\bx)=0~~.
\eeq

When $g_m$ is small we can compute $\bar{u}_l$ to lowest order
in $g_m$~. If we set 
\beq
Q^g_0(\bx)=Q_0(\bx)+q_1(\bx)+\cdots~~,
\eeq
then we easily find
\beq
q_1(\bx)=-\int d^D\bx' G(\bx,\bx')\tau\kk_{lm}\d'_lQ_0(\bx')g_m~~.
\eeq
The mean velocity is, to $O(g)$, 
\beq
\bar{u}_l=\tau\left(\kk_{lm}-\int d^D\bx d^D\bx'
                     \tau\kk_{ln}\d_n\phi(\bx)G(\bx,\bx')
                                \kk_{rm}\d'_rQ_0(\bx')\right)g_m~~. 
\eeq
From eq(\ref{MEANVEL}) we find
\beq
\bar{u}_l=\tau\left(\kk_{lm}-\int d^D\bx d^D\bx'
                              u_l(\bx)G(\bx,\bx')u_m(\bx')Q_0(\bx')\right)g_m~~.
\eeq

This is also the result we obtained in a previous paper. It shows that the 
drift tensor is renormalized in the same way as the diffusivity tensor
if we start from a situation where they are proportional at the microscopic level.
Because this proportionality of the effective drift and diffusivity tensors 
holds sample by sample it also holds after averaging over the ensemble of 
random media. The precise nature of the statistics of the velocity field are 
not important for this result. There is no requirement for Gaussian statistics,
for example. 

The result will also hold in the limit in which sample size 
becomes infinite, assuming the absence of very long range correlations. 
Consequently the result 
guarantees the Ward identity that was discussed in previous work \cite{USDIF2,USDIF3,USDIF4}.
It remains remarkable that the direct proof of this identity,
even in perturbation theory, is so unobvious that it has only
been achieved at the two-loop level. It would be very interesting to
obtain a proof of the Ward identity directly in the infinite medium
to all orders in perturbation theory.

The results confirm those of Kravtsov, Lerner and Yudson \cite{RUSS1,RUSS2} in the 
case of isotropic diffusion, drift and (Gaussian) statistics.
Our result shows that the only important point, as inferred
by these authors, is the existence of the potential controlled
stationary distribution with a vanishing microcurrent. In fact 
by an appropriate linear transformation on the independent variables
$\bx$, the diffusivity tensor can be rendered isotropic and simultaneously, the
drift tensor rendered diagonal. Aside from the question of the statistics
of the field $\phi(\bx)$ therefore, the more general problem considered here
is mathematically identical to the isotropic case.

\section{Incompressible Flow}

Much work has been done on incompressible flow \cite{kraichnan,PHYTH,itdetal1}. 
The velocity field  
satisfies
\beq
\d_lu_l(\bx)=0~~.
\eeq
The diffusion equation for the probability distribution can
can be written
\beq
\fr{\d}{\d t}P(\bx,t)=(\kk_{ij}\d_j-u_i(\bx))\d_iP(\bx,t)~~.
\eeq
It is then easy to see that the static distribution satisfies
\beq
\d_iQ_0(\bx)=0~~.
\eeq
This implies that $Q_0=1/V$~. The role played by the mean velocity
in this case is rather trivial. It is just the centre of mass motion
of the incompressible fluid. For this reason, as noted in \cite{RUSS1, RUSS2}, 
it is never renormalized by the
fluctuations in the velocity field.

In an appropriate reference frame then, the mean velocity 
is zero. That is
\beq
\int_{\Omega}d^D\bx u_l(\bx)=0~~.
\eeq
From eq(\ref{EFFDIFF0}) it follows that
\beq
\kk^{\rm{e}}_{lm}=\kk_{lm}+\int_{\Omega}d^D\bx d^D\bx'
                       u_l(\bx)G(\bx,\bx')u_m(\bx')/V~~.
\eeq
This is similar to eq(\ref{EFFDIFF1}) but with a change of
sign for the correction to the bare diffusivity.
This implies that the effect of the incompressible flow 
is to enhance the diffusivity. In fact in practical cases
the contribution of turbulent dispersion to the effective
diffusivity dominates that of molecular diffusion by many
orders of magnitude. In this case it is reasonable to
consider a limit in which the molecular diffusivity goes
to zero. This is not possible in the gradient flow
case where the particle will simply become trapped at the first 
stationary point it encounters in the potential.

\section{Large Sample Limit}

For completeness we briefly discuss the large sample limit
which we have assumed is innocuous in the sense that 
appropriate spatial averages are equivalent to ensemble
averages. A simple example is the normalization of the
static probability distribution in the case of gradient flow.
We have
\beq
Q_0(\bx)=Ne^{\tau\phi(\bx)}~~.
\eeq
Consider
\beq
X=\int_{\Omega} d^D\bx e^{\tau\phi(\bx)}~~.
\eeq
The ensemble average, in the case of Gaussian statistics, is
\beq
\la X\ra=\int_{\Omega} d^D\bx \la e^{\tau\phi(\bx)}\ra
                       =\int_{\Omega}d^D\bx e^{\fr{1}{2}\tau^2 \Delta(0)}
                       =Ve^{\fr{1}{2}\tau^2 \Delta(0)}~~,
\eeq 
where $\Delta(\bx-\bx')=\la\phi(\bx)\phi(\bx')\ra$~.
We also have
\beq
\la X^2\ra=\int_{\Omega} d^D\bx d^D\bx'
              \la e^{\tau\phi(\bx)+\tau\phi(\bx')}\ra
                        =\int_{\Omega} d^D\bx d^D\bx' e^{\tau^2\Delta(0)+
\tau^2\Delta(\bx-\bx')}~~.
\eeq
The variance of $X$ is therefore
\beq
\sigma^2(X)=\int_{\Omega} d^D\bx d^D\bx' e^{\tau^2\Delta(0)}
\left[e^{\tau^2\Delta(\bx-\bx')}-1\right]
              =Ve^{\tau^2\Delta(0)}\int_{\Omega} d^D\bx \left[e^{\tau^2
\Delta(\bx)}-1\right]~~.
\eeq
It follows that in the limit of large samples in which $V\rightarrow\infty$,
both the mean and variance of $X$ are $O(V)$~. That is for large $V$ we 
may use the result
\beq
X=Ve^{\fr{1}{2}\tau^2\Delta(0)}\left(1+\xi \right)~~,
\eeq
where $\xi$ has zero mean and $\sigma^2(\xi)=O(1/V)$, which suggests
that $\xi$ can be ignored in the large sample limit. This tells us that
in this limit we can use the normalization
\beq
N=e^{-\fr{1}{2}\tau^2\Delta(0)}/V~~.
\eeq

A careful analysis of perturbation theory suggests that similar
results hold for other appropriate quantities. In particular
we expect for large samples, 
\beq
\mu(\bk)=\la\mu(\bk)\ra+O(1/\sqrt{V})~~.
\eeq
Which implies that the infinite medium effective probability distribution
is
\beq
{\cal{P}}(\bx,t)=\int \fr{d^D\bk}{(2\pi)^D} \la e^{i\bk.\bx-\mu(\bk)t}\ra
             =\int  \fr{d^D\bk}{(2\pi)^D} e^{i\bk.\bx-\la\mu(\bk)\ra t}~~.
\eeq
This result was assumed in our discussion above.

\section{Spatially Varying Diffusivity}

So far we have applied the Derrida argument to models for
which the diffusivity tensor is constant in space. Clearly
the argument can be generalised to accommodate spatial fluctuations
in the diffusivity. We therefore make the replacement
\beq
\kk_{ij}\rightarrow \kk_{ij}(\bx)~~,
\eeq
where
\beq
\kk_{ij}(\bx+\bl)=\kk_{ij}(\bx)~~,
\eeq
and $\bl$ is an element of the Bravais lattice. Of course we assume that
the correlation length of the fluctuations in the diffusivity tensor
is very much smaller than the size of the cell $\Omega$~. For simplicity
we will assume also that the local drift term vanishes since it will not be needed
for the application we wish to make.
The physical realizations of this problem are in the computation of the bulk
versions of the diffusivity tensor, permeability tensor or permittivity tensor
in the corresponding diffusion, fluid flow or electrodynamical 
problems \cite{jpmega} together with the bulk response to the appropriate
external fields.

If we adopt the diffusion model, the probability density function for a particle in the medium
satisfies the equation
\beq
\fr{\d}{\d t}P(\bx,t)=\d_i\kk_{ij}(\bx)\d_j P(\bx,t)~~.
\eeq
It is immediately obvious that the static solution is
\beq
Q_0(\bx)=1/V~~.
\eeq
In this case the application of the Derrida technique, along the
lines discussed above, leads us to consider the perturbation
analysis of the equation
\beq
H(\bk)Q(\bx)=-\mu(\bk)Q(\bx)~~,
\label{VARDIFDER}
\eeq
where
\beq
H(\bk)= (\d_i + ik_i)\kk_{ij}(\bx)(\d_j + ik_j)~~,
\eeq
and $\mu(\bk)$ is the eigenvalue that vanishes when $\bk=0$~.
In the standard way we write $\mu(\bk)=\mu_1(\bk)+\mu_2(\bk)+\cdots$ and
$Q(\bx)=Q_0(\bx)+Q_1(\bx)+\cdots$~. However in the present case it
is easily established that $\mu_1(\bk)=0$ and that
\beq
Q_1(\bx)=ik_i\int_{\Omega}d^D\bx'G(\bx,\bx')\d'_j\kk_{ij}(\bx')Q_0~~,
\label{GRN1}
\eeq
where the Green's function $G(\bx,\bx')$ satisfies
\beq
\d_i\kk_{ij}(\bx)\d_jG(\bx,\bx')=-\delta(\bx,\bx')~~.
\eeq
  
A quick way to compute $\mu_2(\bk)$ is to integrate eq(\ref{VARDIFDER}) 
over $\Omega$ to obtain
\beq 
\mu(\bk) = -ik_i\int_{\Omega}d^{D}\bx \kk_{ij}(\bx)(\d_j+ik_j) Q(\bx).
\eeq
On picking out the $O(k^2)$ term from this equation, we find
\beq
\mu_2(\bk) = \bar{\kk}_{ij}k_ik_j-ik_i\int_{\Omega}d^{D}\bx \kk_{ij}(\bx)\d_j Q_1(\bx)~~,
\eeq
where $\bar\kk_{ij}$ is the spatial average of the diffusivity tensor,
\beq
\bar\kk_{ij}=\fr{1}{V}\int_{\Omega}d^D\bx\kk_{ij}(\bx)~~.
\eeq
From eq(\ref{GRN1}) we find
\beq
\mu_2(\bk) = \bar{\kk}_{ij}k_ik_j
  +k_ik_j\int_{\Omega}d^{D}\bx d^D\bx'\kk_{il}(\bx)\d_lG(\bx,\bx')\d'_m\kk_{jm}(\bx') ~~,
\eeq
and after integration by parts
\beq
\mu_2(\bk) = \bar{\kk}_{ij}k_ik_j
  -\fr{k_ik_j}{V}\int_{\Omega}d^{D}\bx d^D\bx'
                     [\d_l\kk_{il}(\bx)]G(\bx,\bx')\d'_m\kk_{jm}(\bx') ~~.
\label{eq:difmu}
\eeq

Hence we have obtained an expression for  $\kk_{ij}^{\rm e}$.

We now restrict ourselves to the isotropic case 
where $\kk_{ij}(\bx) = \kk_0e^{\tau\phi(\bx)}\delta_{ij}$. In the literature 
the case of lognormal statistics where $\phi(\bx)$
is a zero mean Gaussian field with two point function $\Delta(|\bx-\bx'|)$ has been widely
studied. It can also be related, by means of the appropriate Green's functions, 
to a diffusion problem with constant
diffusivity $\kk_0$ and drift parameter $\ll_0=\tau\kk_0$~.  The Green's function
for this latter problem, $G^{grad}(\bx,\bx')$, satisfies
\beq
\d_i\kk_0(\d_i-\tau\d_i\phi(\bx))G^{grad}(\bx,\bx')=-\delta(\bx,\bx')~~,
\eeq
whereas $G(\bx,\bx')$ satisfies 
\beq
\d_i\kk_0e^{\tau\phi(\bx)}\d_iG(\bx,\bx')=-\delta(\bx,\bx')~~.
\eeq
It is easy to see now that the Green's function for the original diffusivity problem
is related to that for the gradient problem by
\beq
G(\bx,\bx') = e^{-\tau\phi(\bx)}G^{grad}(\bx,\bx'). 
\eeq
Substituting this into (\ref{eq:difmu}) we get 
\beq
\mu_2(\bk) = \bar{\kk}k^2
  -\fr{k_ik_j}{V}\int_{\Omega}d^{D}\bx d^D\bx'
                     [\d_i\kk_0e^{\tau\phi(\bx)}]e^{-\tau\phi(\bx)}
             G^{grad}(\bx,\bx')\d'_j\kk_0e^{\tau\phi(\bx')} ~~,
\eeq
where 
\beq
\bar{\kk}=\fr{1}{V}\int d^D\bx\kk_0e^{\tau\phi(\bx)}~~.
\eeq
If now we take the ensemble average $\bar{\kk}$ becomes
\beq
\bar{\kk}=\fr{1}{V}\int d^D\bx\kk_0\la e^{\tau\phi(\bx)}\ra=\kk_0e^{\fr{1}{2}\tau^2\Delta(0)}~~,
\eeq
and obtain the result
\beq
\kk^{\rm e}=\kk_0e^{\fr{1}{2}\tau^2\Delta(0)}
               -\fr{1}{D}\tau\kk_0^2\fr{k_ik_j}{V}\int_{\Omega}d^{D}\bx d^D\bx'
                     \la [\d_i\phi(\bx)]G^{grad}(\bx,\bx')\d'_ie^{\tau\phi(\bx')}\ra ~~,
\eeq 
We can compare this with the result of substituting $\kk_{ij}=\kk_0\delta_{ij}$
and $u_i(\bx)=\tau\kk_0\d_i\phi(\bx)$ into eq(\ref{EFFDIFF1}).
\beq
\kk^{\rm{e}}_{lm}=\kk_0\delta_{lm}-\tau^2\kk_0^2\int_{\Omega}d^D\bx d^D\bx'
                        [\d_l\phi(\bx)]G^{grad}(\bx,\bx')[\d_m\phi(\bx')]Q_0(\bx')~~,
\eeq
where $Q_0(\bx)=Ne^{\tau\phi(\bx)}$~. On taking the ensemble average and
using the isotropy we obtain for the effective diffusivity in the gradient flow case
\beq
\kk^{\rm e}_{grad}=\kk_0-\fr{1}{D}\tau^2\kk_0^2\fr{1}{Ve^{\fr{1}{2}\tau^2\Delta(0)}}
                        \int_{\Omega}d^D\bx d^D\bx'
               [\d_m\phi(\bx)]G^{grad}(\bx,\bx')[\d_m\phi(\bx')]e^{\tau\phi(\bx')}~~.
\eeq
The result is that 
\beq
\kk^{\rm e}/\bar \kk = \kk^{\rm e}_{grad}/\kk_0.  
\label{DIFFGRAD}
\eeq
This means that the ratio of the bulk diffusivity to the (ensemble or spatial)
average of the local diffusivity in the fluctuating diffusivity problem
is equal to the ratio of the bulk diffusivity to the local diffusivity 
in the associated gradient drift problem.

For the case of a lognormal diffusivity then we have
\beq
\kk^{\rm e} = e^{\fr{1}{2}\tau^2\DD(0)}\kk^{\rm e}_{grad}.
\eeq
The consequences of this simple result are quite significant. In one dimension
both problems are easily solved exactly and it is known that
\beq
\kk^{\rm e} = \kk_0e^{-\fr{1}{2}\tau^2\DD(0)}
\eeq
and 
\beq 
\kk^{\rm e}_{grad} = \kk_0 e^{-\tau^2\DD(0)}.
\eeq
Hence our result is of course consistent with the well known exact results
in one dimension. 

In two dimensions however the random diffusivity case 
may be solved exactly via duality arguments \cite{Keller, dykhne1, dykhne2},
giving simply $\kk^{\rm e}=1$.
This means that $\kk^{\rm e}_{grad} = e^{-\fr{1}{2}\tau^2\DD(0)}$ 
giving a new {\em exact} result
for the isotropic gradient flow problem in two dimensions. 

In fact,these results in one and two dimensions are
consistent with a renormalisation group calculation in D dimensions 
for the isotropic gradient flow case \cite{USDIF1,dc} that gives 
\beq 
\kk^{\rm e}_{grad} = \kk_0e^{-\tau^2\DD(0)/D}. 
\label{eq:rg} 
\eeq
Although the renormalization group calculation is based on certain plausible
though unproved assumptions, it maintains the equality of the drift and diffusion
renormalization factors and has been accurately verified in three dimension
by numerical simulation over a considerable range in the value of this factor.
It has been verified by analytical calculation to two loop-order in perturbation theory.
The numerical simulation suggests however that it remains accurate beyond this
order.

Eq(\ref{DIFFGRAD}) allows us to relate these 
renormalization group results to the diffusivity problem.
There has been for some time, a speculation in the literature, for example see \cite{Mat}, 
again verified to two-loop order in perturbation theory, 
that in the diffusivity problem
\beq 
\kk^{\rm e} = \kk_0e^{({1\over 2}- {1\over D})\tau^2\DD(0)} 
\label{eq:diff} .
\eeq
Because eq(\ref{DIFFGRAD}) is an exact relation it shows that they stand
or fall together.
Recently \cite{dewit, AbrInd} it has been
shown that, in three dimensions, at third loop order eq(\ref{eq:diff}) breaks 
down and corrections explicitly depending on the spatial structure of the
correlation function appear. This would mean therefore that the 
formula (\ref{eq:rg}) similarly fails at three loop order in three dimensions.
So far it has not proved possible to analyse this problem to three loops directly.
However a detailed analysis relating the two calculations should make further
progress possible in the gradient drift problem. It is worth noting that 
even if a breakdown of the simple formulae is established, the success
of eq(\ref{eq:rg}) in describing the results of numerical simulations
suggests that it is not necessarily a severe one. 

\section{Conclusions}

We have developed the Derrida technique \cite{DERRIDA}, which was originally
introduced for lattice models \cite{jpmega, DERRIDA}, to the continuum case. It shows
very conveniently the relationship between the effective drift 
and diffusivity tensors in general random media. In particular,
for gradient flow models, it makes clear that when the microscopic
drift and diffusivity tensors are proportional to one another
the corresponding effective tensors show precisely the same
proportionality \cite{USDIF1}. This guarantees the validity, under very
general statistical regimes, of a Ward identity relating
the effective propagator and vertex in wave vector space, derived
up to the two-loop level in previous papers \cite{USDIF2,USDIF3,USDIF4}. It ties in with the
ideas behind the Einstein relation of the drift \cite{RUSS1,RUSS2}
and diffusivity by a temperature factor. Here the role of temperature is
played by the inverse of the proportionality factor $\tau$~.
Of course this may have no direct connection with the real temperature 
of the sample. 

Applied to incompressible flow the method recovers immediately the
non-renormalization of the mean flow by fluctuations in the velocity
field. This is an obvious result on physical grounds and confirms
previous work on the renormalization group approach to anomalous diffusion \cite{RUSS1,RUSS2}.

The method can also be applied to the problem of a random medium with a spatially 
fluctuating diffusivity. It reveals for the isotropic log-normal case
the relationship of the bulk diffusivity with that for a related gradient flow
model. This relation is confirmed in the case of one and two dimensional models
and provides a constraint on the predictions of models in higher dimensions.
The violation of simple results for the bulk diffusivity in one model being tied to
possible failure of simple results in the other.

In principle the formulae derived from the Derrida method also allows the calculation of 
higher order terms that control the approach to asymptotic behaviour
of the various moments of the probability distribution. However the exploitation of
such results would require a proper elucidation of the commutation of the
large sample and large time limits. This is an important mathematical problem
that remains to be investigated.

\end{document}